\documentclass[prd,aps, preprintnumbers, showpacs,twocolumn, nofootinbib,superscriptaddress,notitlepage]{revtex4-1}
\usepackage[normalem]{ulem}
\usepackage{epsfig}
\usepackage{amsfonts}
\usepackage{amsmath}
\usepackage{slashed}
\usepackage{graphicx}
\usepackage{color}
\usepackage{mathtools}
\usepackage{subfigure}
\usepackage{graphicx}

\usepackage{stmaryrd}
\usepackage{amssymb,psfrag}
\usepackage[normalem]{ulem}
\usepackage{caption}

\newcommand{\beq}{\begin{eqnarray}}
\newcommand{\eeq}{\end{eqnarray}}

\newcommand{\nn}{\nonumber}

\def\slash#1{#1 \hskip-0.45em /}

\begin{document}

\title{Factorization Formula Connecting the Shape Functions of Heavy Meson in QCD and Heavy Quark Effective Theory}

\author{Wei Wang}
\email{wei.wang@sjtu.edu.cn}
\affiliation{INPAC, Key Laboratory for Particle Astrophysics and Cosmology (MOE), Shanghai Key Laboratory for Particle Physics and Cosmology,
School of Physics and Astronomy, Shanghai Jiao Tong University, Shanghai 200240, China}

\author{Ji Xu}
\email{xuji\_phy@zzu.edu.cn}
\affiliation{School of Nuclear Science and Technology, Lanzhou University, Lanzhou 730000, China}

\author{Qi-An Zhang}
\email{zhangqa@buaa.edu.cn}
\affiliation{School of Physics, Beihang University, Beijing 102206, China}

\author{Shuai Zhao}
\email{zhaos@tju.edu.cn}
\affiliation{Department of Physics, School of Science, Tianjin University, Tianjin 300350, China}

%\date{\today}
\begin{abstract}
  The shape function of $B$-meson defined in heavy quark effective theory (HQET) plays a crucial role in the analysis of inclusive $B$ decays, and constitutes one of the dominant uncertainties in the determination of CKM matrix element $|V_{ub}|$. On the other hand, the conventional heavy meson shape function defined in QCD is also phenomenologically important and includes shortdistance physics at energy scales of the heavy quark mass. In this work, we derived a factorization formula relating these two kinds of shape functions, which can be invoked to fully disentangle the effects from disparate scales $m_b$ and $\Lambda_{\textrm{QCD}}$, particularly to facilitate the resummation of logarithms $\ln m_b/\Lambda_{\textrm{QCD}}$. In addition, this factorization constitutes an essential component of the recently developed two-step factorization scheme, enabling lattice QCD calculations of lightcone quantities of heavy meson. The results presented here pave the way for first-principles nonperturbative predictions of shape function in the near future.
\end{abstract}

\maketitle

%%%%%%%%%%%%%%%%%%%%%	
\section{Introduction}
\label{Introduction}
%%%%%%%%%%%%%%%%%%%%%
The standard model offers an exceptionally successful description of particle physics, but some key questions remain, especially regarding the flavor physics. One of these is the precise determination of the magnitude of Cabbibo-Kobayashi-Maskawa (CKM) matrix element $|V_{ub}|$. In $b$-flavor sector, the renowned ``$|V_{ub}|$ puzzle'' refers to a significant discrepancy between the measurements obtained from inclusive and exclusive $B$-meson decays \cite{ParticleDataGroup:2024cfk,HFLAV:2022esi,Cao:2023rku}. The determination of $|V_{ub}|$ from exclusive decays predominantly relies on $B \to \pi \ell \nu$, whose theoretical precision is currently limited by our understanding of the $B$-meson light-cone distribution amplitude (LCDA) \cite{Grozin:1996pq}. The extraction of $|V_{ub}|$ from inclusive decays depends critically on our knowledge of the $B$-meson shape function \cite{Neubert:1993ch}. There has been considerable research on the LCDA \cite{Braun:2003wx,Lange:2003ff,Kawamura:2001jm,Ma:2005vv,Li:2006jb,Bell:2013tfa,Braun:2019wyx,Beneke:2018wjp,Gao:2019lta,Cui:2022zwm,Gao:2024vql,Chai:2022ptk,Shi:2024laj}, but investigations into the shape function are comparatively less \cite{Neubert:1993um,Bosch:2004th,Bauer:2003pi,Aglietti:1999ur,Balzereit:1998yf,Bauer:2000ew,Ligeti:2008ac}.

The inclusive decays of $B$-meson, such as the semileptonic process $B \to X_u \ell \nu$ and the radiative process $B \to X_s \gamma$, are of great importance to address the ``$|V_{ub}|$ puzzle'' and play a crucial role in the search for New Physics at $B$ factories. The shape function of $B$-meson, which is an inherent part of factorization theorems for these processes, offers profound insights into the nonperturbative realm of QCD. This physical quantity is crucial since an understanding of the connection between quark and hadron properties is a prerequisite for accurately determining the CKM matrix element. The shape function is defined within the framework of heavy-quark effective theory (HQET), which describes the lightcone residual momentum distribution of the heavy quark inside the $B$-meson. Its physical significance in $B$-meson is similar to that of parton distribution functions (PDFs) in nucleons. Alternatively, analogous to the motion of nucleons within an atomic nucleus, it describes the ``Ferimi motion'' of the $b$-quark inside a $B$-meson.

The shape function of heavy meson lies at the heart of analyses on inclusive decays, as it is universal and encapsulates all the nonperturbative effects. Besides, moments of the $B$-meson shape function offer a natural framework to define heavy-quark parameters such as the $b$-quark mass and kinetic energy \cite{Neubert:2004sp}; in certain models, the shape function of $B$-meson can be directly connected to its LCDAs \cite{LeYaouanc:2007qse}; however, the precision of relevant theoretical predictions is currently limited by the uncertainties of the shape function. Although the studies of shape function including its expansion in a set of orthonormal basis, renormalization group equation (RGE), anomalous dimension and etc., have been carried out \cite{Bosch:2004th,Balzereit:1998yf,Ligeti:2008ac}, a reliable and comprehensive description of its full distribution remains unavailable. Most of the studies on shape function are model-dependent, and the selection of specific model inevitably introduces uncertainties and biases \cite{Lee:2005pwa}. The endeavor to derive $B$-meson shape function from first principles is highly challenging for several reasons. Firstly, this quantity is defined on the lightcone, rendering it difficult to directly handle in lattice QCD. Secondly, the definition of shape function incorporates the HQET field, introducing further complications in lattice simulation. Furthermore, the presence of cusp divergences in $\alpha_s$ correction renders the operator product expansion (OPE) for the heavy meson shape function unsuitable.

The shape function is commonly defined in HQET, i.e., in the infinite-quark mass limit, yet it is also feasible to be defined in the context of QCD. However, for a heavy meson, the shape function defined in QCD encapsulates the hadronic physics at two distinct scales ($m_b$ and $\Lambda_{\textrm{QCD}}$), with the latter expected to manifest infrared behaviors identical to those of the shape function defined in HQET. As for the former scale $m_b$, it can be calculated using perturbation theory. Therefore, a factorization formula connecting these two types of $B$-meson shape functions can be established to fully disentangle the effects from different scales and resum potentially large logarithms between $\Lambda_{\textrm{QCD}}$ and $m_b$. This factorization merges the virtues of both shape functions. Given that much phenomenological knowledge on the shape function in HQET has been garnered from inclusive $B$ decays, we can utilize this factorization to ascertain the starting point of heavy quark mass evolution in full QCD. More generally, such a factorization would provide us with some profound insights into the structure of heavy mesons and is thus interesting in its own right.

In view of lattice QCD applications, this factorization holds considerable significance as well. The authors in \cite{Han:2024min,LatticeParton:2024zko,Han:2025odf} introduced an innovative approach for simulating the $B$-meson LCDA on the lattice, which can be summarized as a two-step factorization scheme. In the first step, the LCDA defined in QCD are derived via a lattice-computable equal-time correlated matrix element in the frame of large momentum effective theory (LaMET) \cite{Ji:2013dva,Ji:2014gla,Cichy:2018mum,Wang:2019msf}. In the second step, the QCD LCDA is converted to HQET LCDA through factorization formula established in \cite{Ishaq:2019dst,Beneke:2023nmj}. The preliminary findings in the two-step factorization scheme qualitatively align with phenomenological models, and is consistent with the experimentally constrain from $B\to \gamma \ell \nu_\ell$ \cite{Belle:2018jqd}, suggesting a promising prospect for delivering first-principle predictions for lightcone observables of heavy meson. Therefore, one can also utilize the two-step factorization scheme to simulate the $B$-meson shape function on the lattice. Naturally, the prerequisite is to construct the relationship between the two types of shape functions in HQET and QCD respectively.

The goal of this work is therefore to establish a factorization formula for the $B$-meson shape functions, thereby enhancing theoretical insight into the structure of heavy mesons, as well as advancing the lattice simulations in the near future. Accurate shape function will have a direct impact on the analysis of inclusive $B$ decays, and even reveal indications for new physics beyond the standard model.

The paper is organized as follows. In section \ref{defSFsandFacFormula}, we proceed by reviewing the construction of $B$-meson shape functions defined in HQET and QCD, followed by a discussion on the factorization formula between them. In section \ref{loopCal}, the one-loop corrections of shape functions and matching function are exhibited, which are the main results of this work. Section \ref{numericalAnaly} contains a numerical study of the $B$-meson shape functions. We conclude in section \ref{summary}. The appendix collects technical details and supplementary results.

%%%%%%%%%%%%%%%%%%%
\section{$B$-meson shape functions and the factorization formula}
\label{defSFsandFacFormula}
%%%%%%%%%%%%%%%%%%%
%%%%%%%%%%%%%%%%%%%
\subsection{$B$-meson shape functions defined in HQET and QCD}
%%%%%%%%%%%%%%%%%%%
For inclusive decays of $B$-meson which are dominated by lightlike distances, the underlying hadronic dynamics are described by the shape function in HQET, defined in terms of the forward matrix elements of  non-local operators spread along the lightcone \cite{Neubert:1993um}
\begin{eqnarray}\label{DefishapefunctionHQET}
  && S^{\textrm{HQET}}(\omega,\mu) = \int_{-\infty}^{+\infty} \frac{dt}{2\pi} e^{i \omega v^+ t} \nn\\
  &&\quad \times \frac{\langle B(v) | \bar h_v(0) \, W(0,tn_+) \, h_v(tn_+) | B(v) \rangle}{\langle B(v) | \bar h_v(0) \, h_v(0) | B(v) \rangle} \,.
\end{eqnarray}
The shape function describes the probability to find a $b$-quark with residual momentum $\omega$ inside the $B$-meson. Here the heavy quark field $h_v$ is defined in HQET with velocity $v_\mu$ satisfying $v^2=1$; $W(0,tn_+)=\textrm{P}\,\{\textrm{exp} [ -i g_s \int_0^{tn_+} ds \, n_+\!\cdot\!A(sn_+) ]\}$ is Wilson line connecting these two heavy quark fields that ensures gauge invariance; $|B(v)\rangle$ is the $B$-meson state. The notations for lightcone coordinates are
\begin{eqnarray}
	n_{+\mu}=\frac{1}{\sqrt{2}}(1,0,0,1) \,, \quad n_{-\mu}=\frac{1}{\sqrt{2}}(1,0,0,-1) \,,
\end{eqnarray}
and we denote $v_{+}\equiv n_{+} \!\cdot\! v$, $v_{-} \equiv n_{-} \!\cdot\! v$ for convenience.

The shape function defined in Eq.\,(\ref{DefishapefunctionHQET}) has support for $-\infty < \omega \leq \bar\Lambda$ with $\bar\Lambda = m_B-m_b$, where $m_b$ is the heavy quark pole mass. The asymptotic behavior of shape function can be calculated perturbatively \cite{Bosch:2004th}, which reveals that it is not positive definite, but acquires a negative radiative tail at large values of $|\omega|$.

The explicit definition for the shape function defined in QCD is
\begin{eqnarray}\label{DefishapefunctionQCD}
 && S^{\textrm{QCD}}(x,\mu) = \int_{-\infty}^{+\infty} \frac{dz^-}{2\pi} e^{-i x p_B^+ z^-} \nn\\
 &&\quad \times \frac{\langle B(p_B) | \bar b(0) \, \Gamma \, W(0,z) \, b(z) | B(p_B) \rangle}{\langle B(p_B) | \bar b(0) \, \Gamma b(0) | B(p_B) \rangle} \,.
\end{eqnarray}
Here $z^2=0$ denotes the lightcone separation and $0\leq x \leq 1$ signifies the lightcone momentum fraction of the $b$-quark inside the hadron; $p_B$ denotes the momentum of the $B$-meson. We select the Dirac matrix $\Gamma=\slash{n}_+$, which ensures that the above definition aligns with that of the quark PDFs of a nucleon \cite{Izubuchi:2018srq}. However, in the case of $B$-meson, the $S^{\textrm{QCD}}(x,\mu)$ involves two distinct scales, $m_b$ and $\Lambda_{\textrm{QCD}}$, of which the former should be amenable to perturbative treatment.

Intuitively, one expects the shape function to be highly asymmetric with the heavy $b$-quark typically carrying most of the lightcone momentum fraction in the heavy meson, which is clearly illustrated in Eq.\,(\ref{treecoeH}) below. It is convenient to separate the perturbative heavy quark mass scale $m_b$ from the nonperturbative hadronic physics in the parameter
\begin{eqnarray}
  \lambda = \frac{\Lambda_{\textrm{QCD}}}{m_b} \ll 1 \,.
\end{eqnarray}
It is therefore necessary to consider the shape function separately in different regions. The region where $x \sim  1-\lambda $ is referred to as the ``peak region'', which is characterized by large momentum fractions of the heavy quark. While the region $x \sim 0 $ is termed the ``tail region'', in which the behavior of $S^{\textrm{QCD}}(x,\mu)$ can be computed perturbatively, and its result will be presented in Eq.\,(\ref{loopcoeHtail}) below.

%%%%%%%%%%%%%%%%%%%
\subsection{Factorization formula}
%%%%%%%%%%%%%%%%%%%
The shape functions in Eq.\,(\ref{DefishapefunctionHQET}) and Eq.\,(\ref{DefishapefunctionQCD}) exhibit distinct ultraviolet behaviors, as reflected in their differing scale evolution kernels. Nevertheless, it is crucial to observe that these two objects share precisely the same infrared behavior, as HQET is constructed to replicate the (infrared) IR behavior of QCD. In addition, the role of $m_b$ differs between these two theories, $S^{\textrm{QCD}}(x,\mu)$ incorporates dependence on the heavy quark mass, whereas $S^{\textrm{HQET}}(\omega,\mu)$ does not at all. This necessitates a refactorization procedure,
\begin{eqnarray}\label{fac1}
  S^{\textrm{QCD}}(x,\mu) \!=\!
  \begin{cases}
  Z_{\textrm{peak}}(x, \omega, \mu) \otimes S^{\textrm{HQET}}(\omega,\mu) \,, & x \!\sim\! 1 \!-\! \lambda \\ \\
  Z_{\textrm{tail}}(x,\mu) \,, & x \sim 0
  \end{cases}
\end{eqnarray}
where $\otimes$ denotes a convolution in the variable $\omega$. The determination of $Z$ function above can be reliably achieved through the standard perturbative matching procedure. Since this function is insensitive to the IR physics, the $B$-meson state can be safely replaced with a free $b \bar q$ quark state, allowing $S^{\textrm{QCD}}(x,\mu)$ and $S^{\textrm{HQET}}(\omega,\mu)$ to be computed in perturbation theory:
\begin{subequations}\label{expansions}
\begin{eqnarray}
  S^{\textrm{QCD}}(x,\mu) &=& S^{\textrm{QCD}(0)}(x,\mu) \nn\\
  && \!\!\!\!\! + \frac{\alpha_s C_F}{2\pi} S^{\textrm{QCD}(1)}(x,\mu) +\mathcal{O}(\alpha_s^2) \,,\\\nn\\
  S^{\textrm{HQET}}(\omega,\mu) &=& S^{\textrm{HQET}(0)}(\omega,\mu) \nn\\
  && \!\!\!\!\! + \frac{\alpha_s C_F}{2\pi} S^{\textrm{HQET}(1)}(\omega,\mu) +\mathcal{O}(\alpha_s^2) \,.
\end{eqnarray}
\end{subequations}
One is then able to solve Eq.\,(\ref{expansions}) to deduce the $Z$ function, order by order in $\alpha_s$,
\begin{subequations}\label{Zexpansions}
\begin{eqnarray}
  Z_{\textrm{peak}}(x, \omega, \mu) &=& Z_{\textrm{peak}}^{(0)}(x, \omega, \mu) \nn\\
  && \!\!\!\!\! + \frac{\alpha_s C_F}{2\pi}Z_{\textrm{peak}}^{(1)}(x, \omega, \mu) +\mathcal{O}(\alpha_s^2) \,,\\\nn\\
  Z_{\textrm{tail}}(x, \mu) &=& Z_{\textrm{tail}}^{(0)}(x, \omega, \mu) \nn\\
  && \!\!\!\!\! + \frac{\alpha_s C_F}{2\pi}Z_{\textrm{tail}}^{(1)}(x, \mu) +\mathcal{O}(\alpha_s^2) \,.
\end{eqnarray}
\end{subequations}
Identifying momentums of the $B$-meson and $b$-quark as $p_B^+ = m_B v^+$, $p_b^+ = m_b v^+ + k^+$. At tree level, Eq.\,(\ref{fac1}) reduces to
\begin{eqnarray}
  \delta (xm_B v^+ \!\!-\! m_b v^+ \!\!-\! k^+) \!\!=\!\! \int_{-\infty}^{\bar\Lambda} \!\!\! d\omega \, Z_{\textrm{peak}}^{(0)}(x, \omega, \mu) \delta(\omega v^+ \!\!-\! k^+) \,.\nn\\
\end{eqnarray}
This  leads to the tree level result of matching function,
\begin{subequations}\label{treecoeH}
\begin{eqnarray}
  Z_{\textrm{peak}}^{(0)}(x, \omega, \mu) &=& \delta( \omega v^+ +m_b v^+ -xm_B v^+ ) \,, \\
  Z_{\textrm{tail}}^{(0)}(x, \omega, \mu) &=& 0 \,.
\end{eqnarray}
\end{subequations}
Contributions from the tail region arise only at the $\alpha_s$ order. The $\delta$ function guarantees that $\omega$ with support $(-\infty, \bar\Lambda)$ is monotonically mapped onto $x$ with support $(0, 1)$. Please note that the typical momentum of $\omega$ is soft, therefore the shape function defined in QCD has a typical support in a small region close to the endpoint,
\begin{eqnarray}
  x = \frac{\omega +m_b }{m_B } \sim 1 \,.
\end{eqnarray}

%%%%%%%%%%%%%%%%%%%
\section{Factorization formula at next-to-leading order}
\label{loopCal}
%%%%%%%%%%%%%%%%%%%
%%%%%%%%%%%%%%%%%%%
\subsection{Preliminaries}
%%%%%%%%%%%%%%%%%%%
We now turn to a study of radiative corrections to the shape functions. By solving Eq.\,(\ref{fac1}) at next-to-leading order in the peak region,
\begin{eqnarray}\label{matchingoneloop}
  S^{\textrm{QCD}(1)}(x,\mu) &=& \left. S^{\textrm{HQET}(1)}(\omega,\mu) \right|_{\omega v^+ \to x m_B v^+ -m_b v^+} \nn\\
  &&+ \left. Z_{\textrm{peak}}^{(1)}(x, \omega, \mu) \right|_{\omega v^+ \to k^+} \,.
\end{eqnarray}
The above expression can assist us in deriving the perturbative expressions for matching function $Z$ at order $\alpha_s$.

Before delving into the details, we first outline several key points to help clarify the calculated results presented below.
\begin{itemize}
  \item The shape function focus on the residue momentum of $b$-quark, which has a typical scale of $\Lambda_{\textrm{QCD}}$. This necessitates a region-separated factorization formula, as in the case of tree-level.

  \item In the subsequent calculations, we will work in the space-time dimension $d=4-2\epsilon$ and use $\overline{\textrm{MS}}$ scheme to renormalize the ultraviolet (UV) divergences. In order to identify the relevant IR singularities, we keep $v \cdot k$ non-zero in this work.

  \item It is convenient to regulate the singularity that arises when $\omega v^+ \to k^+$ or $x m_B v^+ \to m_b v^+ +k^+$ by a distribution function. In essence, the so-called star distributions and $\mu$ distribution introduced in \,\cite{Bauer:2003pi,Bosch:2004th} are equivalent. In this work, we employ the modified plus distribution defined in \,\cite{Wang:2019msf},
      \begin{eqnarray}\label{def_plus}
        F\left(\omega, k^{+}\right)=\left[F\left(\omega, k^{+}\right)\right]_{\oplus}+\delta\left(\omega-k^{+}\right) \int_{0}^{\Lambda} d t \, F(\omega, t) \,.\nn\\
      \end{eqnarray}
      Note that the integral range of this plus distribution is restricted to a finite interval  $[0,\Lambda]$.

  \item A model-independent description of the asymptotic behavior of the shape function $S^{\textrm{HQET}}(\omega,\mu)$ defined in HQET for large values of $|\omega|$ has been displayed \cite{Bosch:2004th}, so we omit further discussion on this issue here.
\end{itemize}

\begin{widetext}
%%%%%%%%%%%%%%%%%%%
\subsection{One-loop results for the shape function $S^{\textrm{QCD}}(x,\mu)$ in QCD}
%%%%%%%%%%%%%%%%%%%
The one-loop Feynman diagrams in Fig.\,\ref{All-in} consist of the heavy-quark sail graph, box graph, and local vertex graph. We will list the results of these graphs one by one.
\begin{figure}[htbp]
  \centering
  \includegraphics[width=0.75\columnwidth]{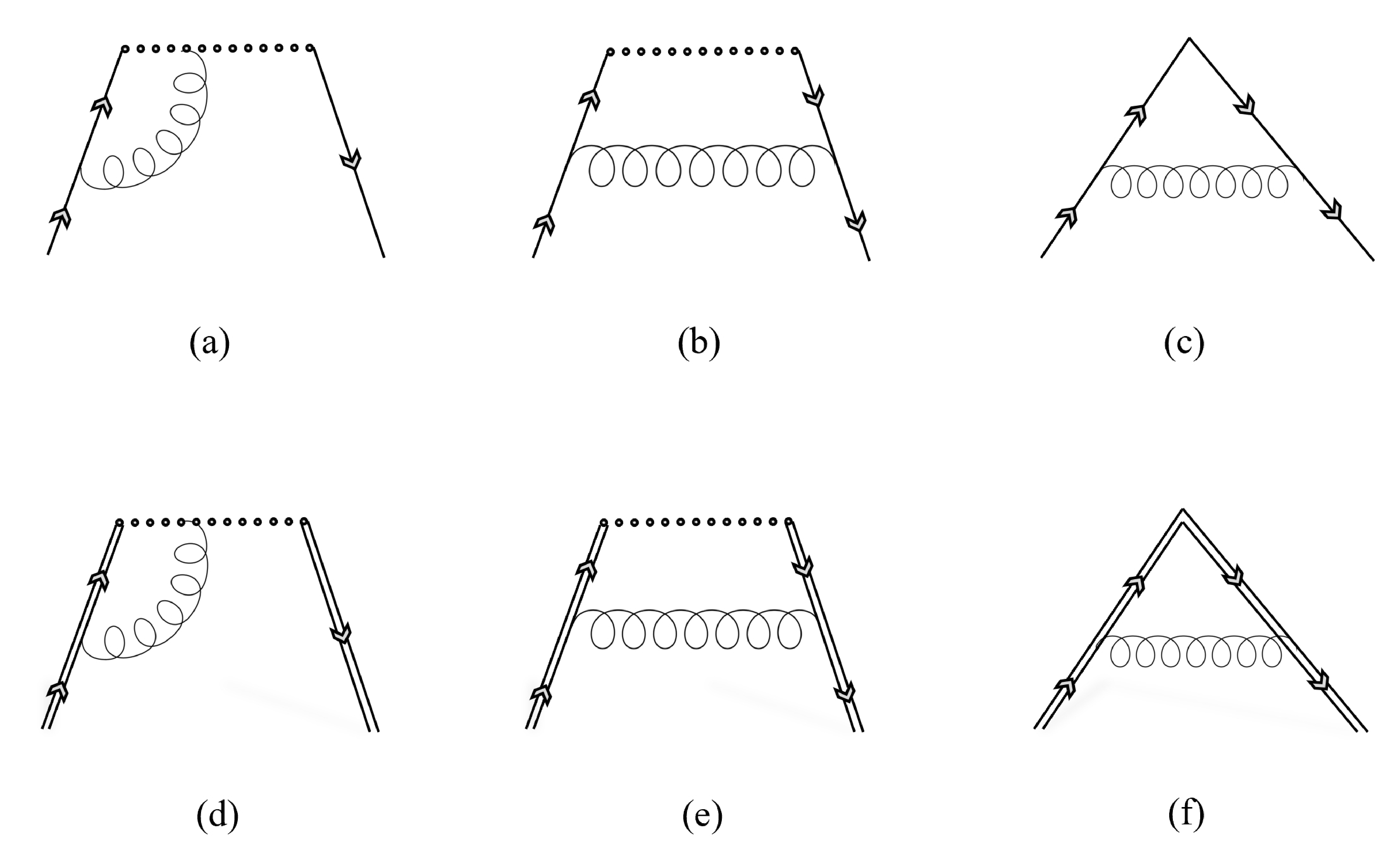}
  \caption{Feynman diagrams for the $B$-meson shape functions in QCD (top row) and HQET (bottom row). The HQET $b$-quark is represented by a double solid line, while the Wilson line appears as a dotted line.}
  \label{All-in}
\end{figure}
The graphs in the top row correspond to the shape function defined in QCD, we start with the heavy-quark sail graph in Fig.\,\ref{All-in}(a),
\begin{subequations}\label{SQCDa}
\begin{eqnarray}
  && \left. S^{\textrm{QCD}(a)}(x,\mu) \right|_{\textrm{peak}} = \bigg[ \frac{2}{m_b v^+ \! -x m_B v^+ \! +k^+} \, \ln\frac{v^{+2}\mu^2}{(m_b v^+ \! -x m_B v^+ \! +k^+)(m_b v^+ \! -x m_B v^+ \! +k^+ \! -2v^+ v\cdot k)}  \nn\\
  && \qquad \times \theta(m_b v^+ \! -x m_B v^+ \! +k^+)\bigg]_\oplus \nn\\
  && \qquad +\bigg[ \int_{xm_B v^+ + k^+}^{\Lambda} dt \, \frac{2}{m_b v^+ \! -xm_B v^+ \! +t} \, \ln\frac{v^{+2}\mu^2}{(m_b v^+ \! -xm_B v^+ \! +t)(m_b v^+ \! -xm_B v^+ \! +t -2v^+ v\cdot k)} \bigg] \nn\\
  && \qquad \times \delta(x m_B v^+ \! -m_b v^+ \! -k^+) \,, \\\nn\\
  %%%%%%
  && \left. S^{\textrm{QCD}(a)}(x,\mu) \right|_{\textrm{tail}} = \frac{1}{m_b v^+} \frac{2x}{1-x} \left[ \ln\frac{\mu^2}{(1-x)^2 m_b^2} \right] \theta(1-x) \,.
\end{eqnarray}
\end{subequations}
The above expressions categorize the results by different regions. The presence of $v \cdot k$ in the denominator serves as an IR regulator, ensuring the integration remains well-defined when $x m_B v^+ \to m_b v^+ +k^+$.

The results of box graph in Fig.\,\ref{All-in}(b) read
\begin{subequations}\label{SQCDb}
\begin{eqnarray}
  && \left. S^{\textrm{QCD}(b)}(x,\mu) \right|_{\textrm{peak}} = \left[ -\frac{2}{k^+ +m_b v^+ -xm_B v^+ -2v^+ v\cdot k} \theta(m_b v^+ -xm_B v^+ +k^+) \right]_\oplus \nn\\
  && \qquad - \left[ \int_{xm_B v^+ -m_b v^+}^{\Lambda} dt \, \frac{2}{t+m_b v^+ -xm_B v^+ -2v^+ v\cdot k} \right] \delta(k^+ +m_b v^+ -xm_B v^+) \,, \\\nn\\
  %%%%%%
  && \left. S^{\textrm{QCD}(b)}(x,\mu) \right|_{\textrm{tail}} = \frac{1}{m_b v^+} \left[ -\frac{1+x^2}{1-x} +(1-x)\ln\frac{\mu^2}{(1-x)^2 m_b^2} \right] \theta(1-x) \,.
\end{eqnarray}
\end{subequations}
The one-loop result of local vertex graph in Fig.\,\ref{All-in}(c) is
\begin{subequations}\label{SQCDc}
\begin{eqnarray}
  && \left. S^{\textrm{QCD}(c)}(\mu) \right|_{\textrm{peak}} = 2-\ln\frac{m_b^3}{(-2v\!\cdot\!k)^2\mu}  \,, \\\nn\\
  %%%%%%
  && \left. S^{\textrm{QCD}(c)}(\mu) \right|_{\textrm{tail}} = 0 \,.
\end{eqnarray}
\end{subequations}
Eqs.\,(\ref{SQCDa})-(\ref{SQCDc}) constitute the one-loop results for the renormalized shape function $S^{\textrm{QCD}}(x,\mu)$ in QCD, details on these graphs regarding the full $\epsilon$ dependence and the separation of regions are provided in Appendix \ref{AppA}.1. The IR regulator $v\cdot k$ in the peak region will cancel out during the matching procedure in Eq.\,(\ref{matchingoneloop}).

%%%%%%%%%%%%%%%%%%%
\subsection{One-loop results for the shape function $S^{\textrm{HQET}}(\omega,\mu)$ in HQET}
%%%%%%%%%%%%%%%%%%%
The asymptotic behavior of the shape function $S^{\textrm{HQET}}(\omega,\mu)$ in large $|\omega|$ (tail region) has been calculated perturbatively in \cite{Bosch:2004th}, we refrain from repeating those results here. We focus our presentation on the factorization formula between $S^{\textrm{QCD}}(x,\mu)$ and $S^{\textrm{HQET}}(\omega,\mu)$ in the peak region. The graphs in the bottom row in Fig.\,\ref{All-in} correspond to the shape function defined in HQET, the result of heavy-quark sail graph in Fig.\,\ref{All-in}(d) is
\begin{eqnarray}\label{HQETd}
  && S^{\textrm{HQET}(d)}(\omega,\mu) = \left[ \left( \frac{2}{k^+ -\omega v^+} \ln\frac{v^{+2}\mu^2}{(k^+ -\omega v^+)(k^+ -\omega v^+ -2v^+ v\!\cdot\!k)} \right) \theta(k^+ -\omega v^+) \right]_\oplus \nn\\
  &&\qquad -\bigg[ \frac{\pi^2}{12} +\frac{1}{2}\ln^2\frac{\omega^2}{\mu^2} -\int_{2\omega v^+}^{\Lambda} dt\, \frac{2}{t-\omega v^+} \ln\frac{v^{+2}\mu^2}{(t -\omega v^+)(t -\omega v^+ -2v^+ v\!\cdot\!k)} \bigg] \delta(\omega v^+ -k^+) \,.
\end{eqnarray}
The result of box graph in Fig.\,\ref{All-in}(e) is
\begin{eqnarray}\label{HQETe}
  && S^{\textrm{HQET}(e)}(\omega,\mu) = -\left[ \frac{2}{k^+ - \omega v^+ -2v^+ v\!\cdot \!k} \theta(k^+ -\omega v^+ ) \right]_\oplus  -\left[ \int_{\omega v^+}^{\Lambda} dt \, \frac{2}{t-\omega v^+ -2 v^+ v\!\cdot \! k} \right] \delta(\omega v^+ -k^+) \,.
\end{eqnarray}
The following result presents the local vertex graph in Fig.\,\ref{All-in}(f)
\begin{eqnarray}\label{HQETf}
 && S^{\textrm{HQET}(f)}(\mu) = -\ln\frac{\mu^2}{(-2 v\!\cdot\! k)^2} \,.
\end{eqnarray}
Eqs.\,(\ref{HQETd})-(\ref{HQETf}) constitute the one-loop results for the renormalized shape function $S^{\textrm{HQET}}(\omega,\mu)$ in HQET. Calculation details are provided in Appendix \ref{AppA}.2. The $v\cdot k$ term serves as a regulator that eliminates potential singularities when $\omega v^+ \to k^+$. This IR dependency would cancel between shape functions defined in QCD and HQET, leaving the matching function free of $v\cdot k$, as it should be.
\end{widetext}

%%%%%%%%%%%%%%%%%%%
\subsection{One-loop results for the matching function}
%%%%%%%%%%%%%%%%%%%
With the one-loop corrections for shape functions calculated above, we now proceed to determine the perturbative matching function entering the factorization formula. We begin with the matching function in the tail region. As seen from Eq.\,(\ref{treecoeH}b), it is power suppressed. Substituting Eq.\,(\ref{Zexpansions}b) into Eq.\,(\ref{fac1}) and combining with the results for $\left. S^{\textrm{QCD}(1)}(x,\mu) \right|_{\textrm{tail}}$ from Eqs.\,(\ref{SQCDa}b)-(\ref{SQCDc}b), the matching function $Z_{\textrm{tail}}(x, \mu)$ at one-loop can be straightforwardly written down:
\begin{eqnarray}\label{loopcoeHtail}
  Z_{\textrm{tail}}^{(1)}(x, \mu) \!=\! \frac{1}{m_b v^+} \frac{1+x^2}{1-x} \! \left[ -1 \!+\! \ln\frac{\mu^2}{(1-x)^2 m_b^2} \right] .
\end{eqnarray}
The momentum fraction $x\sim 0$ corresponds to large values $|\omega| \gg \Lambda_{\textrm{QCD}}$, in which case the tail of shape function $\left. S^{\textrm{HQET}}(\omega,\mu) \right|_{\textrm{tail}}$ in HQET can be determined perturbatively. Likewise, the tail of shape function $\left. S^{\textrm{QCD}}(x,\mu) \right|_{\textrm{tail}}$ in QCD also admits a perturbative calculation.

We now proceed to the matching in the peak region, characterized by momentum fractions $x \sim 0$, which implies a different counting compared to the tail region. The one-loop matrix element of $S^{\textrm{QCD}}(x,\mu)$ is given by the sum of heavy-quark sail graph in Eq.\,(\ref{SQCDa}a), the box graph in Eq.\,(\ref{SQCDb}a) and the local vertex graph in Eq.\,(\ref{SQCDc}a),
\begin{eqnarray}\label{oneloopforSQCD}
  && \left. S^{\textrm{QCD}(1)}(x,\mu) \right|_{\textrm{peak}} \!\!=\! \left. S^{\textrm{QCD}(a)}(x,\mu) \right|_{\textrm{peak}} \!\!+\! \left. S^{\textrm{QCD}(b)}(x,\mu) \right|_{\textrm{peak}} \nn\\
  && \qquad - \left. S^{\textrm{QCD}(c)}(\mu) \right|_{\textrm{peak}} \!\!\times\! \delta(k^+ +m_b v^+ -xm_B v^+) \,.
\end{eqnarray}

Similarly, the one-loop matrix element of $S^{\textrm{HQET}}(\omega,\mu)$ is given by the sum of heavy-quark sail graph in Eq.\,(\ref{HQETd}), the box graph in Eq.\,(\ref{HQETe}) and the local vertex graph in Eq.\,(\ref{HQETf}),
\begin{eqnarray}\label{oneloopforSHQET}
  && \left. S^{\textrm{HQET}(1)}(\omega,\mu) \right|_{\textrm{peak}} \!= S^{\textrm{HQET}(d)}(\omega,\mu) + S^{\textrm{HQET}(e)}(\omega,\mu) \nn\\
  && \qquad - S^{\textrm{HQET}(f)}(\mu) \times \delta(\omega v^+ -k^+) \,.
\end{eqnarray}
By inserting the results for shape functions in Eqs.\,(\ref{oneloopforSQCD})-(\ref{oneloopforSHQET}) into Eq.\,(\ref{matchingoneloop}), we obtain the matching function at order $\alpha_s$ in the peak region,
\begin{eqnarray}\label{loopcoeHpeak}
  && Z_{\textrm{peak}}^{(1)}(x, \omega, \mu) = \left( \frac{1}{2}\ln^2\frac{\mu^2}{m_b^2} -\frac{3}{2}\ln\frac{\mu^2}{m_b^2} +\frac{\pi^2}{12} -2 \right) \nn\\
  &&\qquad \times \delta(xm_B v^+ -m_b v^+ -\omega v^+) \,.
\end{eqnarray}
Note that all the dependence on $x$ and $\omega$ resides in the delta function, which implies the factorization formula reduces from a convolution to a multiplication:
\begin{eqnarray}\label{matchinginpeak}
  && \left. S^{\textrm{QCD}}(x,\mu) \right|_{\textrm{peak}} \!\!=\!\! \left[ 1 \!+\! \frac{\alpha_s C_F}{2\pi}\!\!\left( \frac{1}{2}\ln^2\frac{\mu^2}{m_b^2} \!-\!\frac{3}{2}\ln\frac{\mu^2}{m_b^2} \!+\!\frac{\pi^2}{12} \!-\!2 \!\right) \!\right] \nn\\
  && \qquad \left. \times S^{\textrm{HQET}}(\omega,\mu) \right|_{\textrm{peak}, \,\, \omega v^+ \to xm_B v^+ -m_b v^+}  \,.
\end{eqnarray}
The main result of this work consists of deriving the tree and one-loop results of the matching function in Eq.\,(\ref{treecoeH}), (\ref{loopcoeHtail}) and (\ref{loopcoeHpeak}). Firstly, we note that the $v\cdot k$ terms appearing in the one-loop calculation of the shape functions cancel exactly during matching process, leaving the matching function dependent only on $\mu$ and $m_b$. This implies the validity of the factorization formula, i.e., the two kinds of shape functions exhibit the same infrared structure. Secondly, the plus distributions, irrespective of their specific definition, also cancel out in the matching, yielding remarkably simple form in Eq.\,(\ref{matchinginpeak}). Thirdly, the ``refactorization framework'' developed in this work is in a similar spirit to invoke the refactorization approach to bridge two kinds of $B$-meson LCDAs defined in QCD and HQET, respectively \cite{Ishaq:2019dst,Beneke:2023nmj}.

This factorization formula of $B$-meson shape functions is particularly relevant for lattice QCD implementations. The two-step factorization scheme, originally formulated for simulating $B$-meson LCDA, is equally applicable to shape function calculations. Eq.\,(\ref{fac1}) provides the second step of this scheme, the first step requires constructing the $B$-meson quasishape function within the LaMET framework, which is essentially equivalent to the quark quasiparton distribution function
\begin{eqnarray}\label{DefiquasishapefunctionQCD}
 && \tilde S^{\textrm{QCD}}(x,p_B^z,\mu) = \int_{-\infty}^{+\infty} \frac{dz}{2\pi} e^{i x p_B^z z} \nn\\
 &&\quad \times \frac{\langle B(p_B) | \bar b(0) \, \Gamma \, W(0,z) \, b(z) | B(p_B) \rangle}{\langle B(p_B) | \bar b(0) \, \Gamma b(0) | B(p_B) \rangle} \,,
\end{eqnarray}
where $\Gamma=\gamma^z$, $z_\mu=z n_\mu$, $n_\mu=(0,0,0,1)$. This is a spacelike Euclidean distribution and can be simulated directly on the lattice. $\tilde S^{\textrm{QCD}}(x,p_B^z,\mu)$ can be related to $S^{\textrm{QCD}}(x,\mu)$ through factorization formula (the first step factorization). Extensive expertise has been developed in simulating quasiparton distribution function and quasidistribution amplitude under the LaMET formalism \cite{Xiong:2013bka,Ma:2014jla,Fan:2018dxu,Xu:2018mpf,Liu:2019urm,Ji:2020ect,Xu:2022guw,Xu:2022krn,Hu:2023bba,Hu:2024ebp,Zhao:2019elu,Zhao:2020bsx,Hu:2024mas,LatticeParton:2024vck,Deng:2024dkd,Han:2024cht}, while a crucial consideration here is preserving the heavy meson mass in simulations - a requirement that notably increases the complexity of lattice computations \cite{Wang:2024wwa}. However, this is somewhat off the main topic of this work, we will explicitly illustrate the lattice simulation on the quasishape function as well as the result of $B$-meson shape function in a long write-up.

%%%%%%%%%%%%%%%%%%%
\section{Determining QCD Shape Function from HQET Shape Function}
\label{numericalAnaly}
%%%%%%%%%%%%%%%%%%%
The HQET shape function of $B$-meson is a nonperturbative object, which at present has not been predicted from first principles. However, valuable insights have been gained to construct a realistic model for this quantity \cite{Balzereit:1998yf,Bosch:2004th,Ligeti:2008ac}. In this section, we determine the QCD shape function of $B$-meson in terms of a widely adopted model of HQET shape function at the soft scale $\mu_i=1.5\,\textrm{GeV}$, it is instructive to understand the characteristic feature of QCD shape function and beneficial for future lattice simulations.

The form of the adopted model is
\begin{eqnarray}\label{themodel}
  && \hat{S}^{\textrm{HQET}}(\hat{\omega}, \mu)=\frac{N}{A}\left(\frac{\hat{\omega}}{A}\right)^{b-1} \!\! \exp \left(-b \frac{\hat{\omega}}{A}\right) \nn\\
  &&\qquad -\frac{\alpha_s C_F}{\pi} \frac{\theta(\hat{\omega}-A-\mu / \sqrt{e})}{\hat{\omega}-A} \!\! \left(\! 2 \ln \frac{\hat{\omega}-A}{\mu} \!+\! 1 \!\right) ,
\end{eqnarray}
with the normalization factor $N$ give by
\begin{eqnarray}
  N=\left[1-\frac{\alpha_s C_F}{\pi}\left(\frac{\pi^2}{24}-\frac{1}{4}\right)\right] \frac{b^b}{\Gamma(b)} \,.
\end{eqnarray}
The parameters of this model at the intermediate scale $\mu_i=1.5\,\textrm{GeV}$ are taken as $A=0.685\,\textrm{GeV}$ and $b=2.93$ \cite{Bosch:2004th}. The variable $\hat{\omega}=\bar\Lambda-\omega \geq 0$, and it corresponds to the momentum fraction via the relation $\hat{\omega}=(1-x)m_B$, here we take $m_B=5.28\,\textrm{GeV}$.

\begin{figure}[htbp]
\centering
\includegraphics[width=0.95\columnwidth]{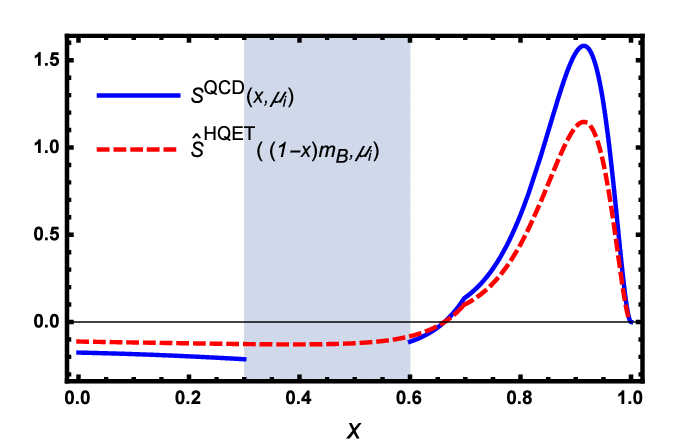}
%\vspace{0.2 cm}
\centering
\caption{The $B$-meson shape function in HQET $\hat{S}^{\textrm{HQET}}(\hat{\omega}, \mu)$ (red dashed line) at the intermediate scale $\mu_i=1.5\,\textrm{GeV}$ is obtained by using the phenomenological model in Eq.\,(\ref{themodel}). The $B$-meson shape function in QCD $S^{\textrm{QCD}}(x,\mu)$ (blue solid line) is derived from the factorization formula in Eq.\,(\ref{fac1}).  The shaded area characterizes the intermediate transition region between the tail and peak regions.
}
\label{diagram12}
\end{figure}

By utilizing the phenomenological model of the $B$-meson shape function given in Eq.\,(\ref{themodel}) and the matching function from Eq.\,(\ref{loopcoeHpeak}), the factorization formula yields the QCD shape function in the peak region, which is illustrated in Fig.\,\ref{diagram12}. For the tail region, we employ the perturbative results from Eq.\,(\ref{loopcoeHtail}), with the numerical inputs set to $m_b=4.65\,\textrm{GeV}$ and $v^+=4$. The shaded area in Fig.\,\ref{diagram12} characterizes the intermediate transition region between the perturbative tail and nonperturbative peak regions. This region presents a unique theoretical challenge: it is not amenable to purely perturbative treatment and requires the inclusion of higher-order $(1-x)$ corrections in the QCD-to-HQET factorization formula. A systematic analysis of this transition region will be addressed in our future work.

%%%%%%%%%%%%%%%%%%%%%%%
%%%%%%%%%%%%%%%%%%%%%%%

%%%%%%%%%%%%%%%%%%%%%%%
\section{Summary and outlook}
\label{summary}
The shape function of $B$-meson has important implications in both precision determinations of CKM matrix elements and enhancing sensitivity to new physics. However, our knowledge on this physical quantity is relatively poor. In this work, we derived a factorization formula Eq.\,(\ref{fac1}) that connects these two kinds of important shape functions of heavy meson defined in QCD and HQET respectively, in the form of a convolution of the latter with a perturbatively calculable matching function, which takes a rather simple multiplication form at the one-loop order considered here. Through this factorization formula and the model of shape function in HQET, we explicitly constructed the $B$-meson shape function in QCD. Performing matching at different heavy quark masses, one can obtain the shape functions from $B$-meson to $D$-meson. For the charmed meson case, power corrections of $\Lambda_{\textrm{QCD}}/m_c$ are sizeable and should be included in the analysis.

The factorization formula presented in this work serves as a prototype for other physical quantities of heavy mesons and heavy baryons. It is a mandatory requirement in the recently developed two-step factorization scheme, which enables lattice simulations on lightcone quantities of heavy meson. The subsequent step urgently required is a dedicated lattice simulation. Achieving this goal demands improved methodologies to control both statistical and systematic uncertainties, along with advances in computing techniques and resources. We will leave this issue for near future studies.

Finally, let us reiterate that the results presented here are valid at leading power in $\Lambda_{\textrm{QCD}}/m_b$ and at one-loop order in $\alpha_s$. While it may be laborious to calculate the power correction and the two-loop order results, our approach allows for their systematic treatment, and we believe that such calculations are essential for achieving theoretically precise predictions of inclusive $B$ decays. By respectively providing crucial nonperturbative inputs for both exclusive and inclusive decay processes (LCDA and shape function), it would lead a way to ultimately resolve the renowned ``$|V_{ub}|$ puzzle''. This would be worth the effort.

\section*{Acknowledgements}
The authors would like to thank Prof. Yong Zhao for fruitful and inspiring discussions. This work is supported in part by National Natural Science Foundation of China under Grant No.  12125503, 12335003, 12475098 and 12105247.

\begin{widetext}
\appendix
%%%%%%%%%%%%%%%%%%%%%%%
\section{One-loop corrections of shape functions}\label{AppA}
%%%%%%%%%%%%%%%%%%%%%%%
%%%%%%%%%%%%%%%%%%%%%%%
\subsection{Shape function $S^{\textrm{QCD}}(x,\mu)$ defined in QCD}
%%%%%%%%%%%%%%%%%%%%%%%
\begin{figure}[htbp]
  \centering
  \includegraphics[width=0.75\columnwidth]{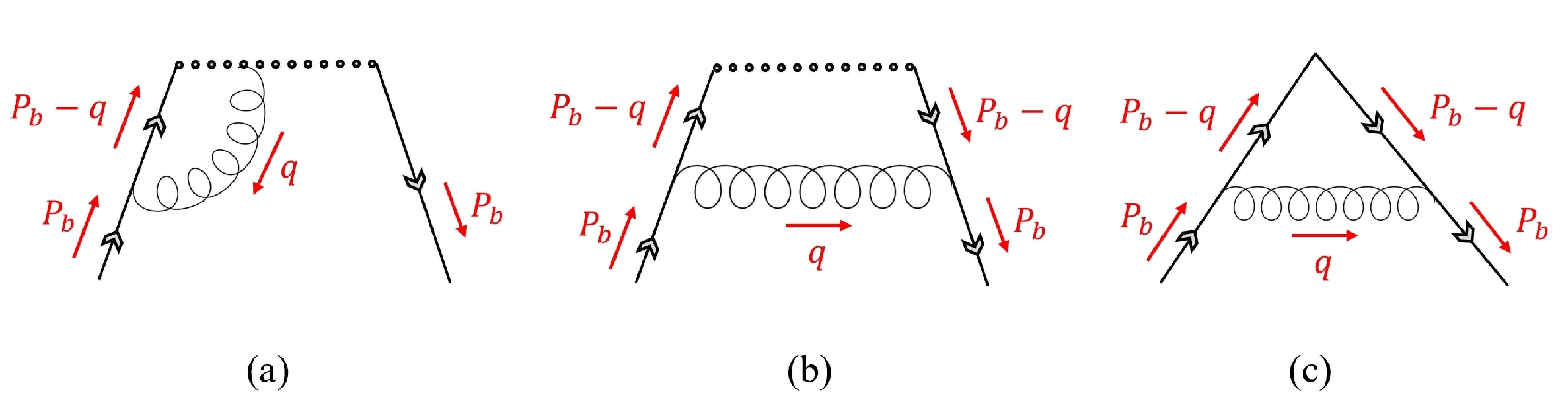}
  \caption{Feynman diagrams for the $B$-meson shape function defined in QCD. The relevant momentum notions are labeled.}
  \label{All-in-QCD}
\end{figure}
We begin with the heavy-quark sail graph, the relevant momentum notions are indicated in Fig.\,\ref{All-in-QCD}(a). The corresponding amplitude is:
\begin{eqnarray}\label{ampQCDa}
  && S^{\textrm{QCD}(a)}(x,\mu) = \frac{2\pi}{\alpha_s C_F} \int \frac{d^d q}{(2\pi)^d} \bar u(p_b) \frac{-i}{q^+ -i\epsilon} (ig_s t^a n_+^\mu \tilde\mu^\epsilon) \slash{n}_+ \frac{-i}{q^2+i\epsilon} \frac{i(\slash{p}_b+\slash{q}+m_b)}{(p_b+q)^2-m_b^2 +i\epsilon} (ig_s t^a \gamma_\mu \tilde\mu^\epsilon) u(p_b) \nn\\
  &&\qquad \times \frac{1}{p_b^+} \left[ \delta(xp_B^+ -p_b^+ -q^+) - \delta(x p_B^+ -p_b^+) \right] \nn\\
  &&\qquad =  - \int_0^{p_b^+}  dq^+ \frac{2(q^+ - p_b^+)}{q^+ p_b^+} \left[ \frac{1}{\epsilon_{\textrm{UV}}} +\ln\frac{v^+(m_b v^+ +k^+)\mu^2}{q^+(2k^- q^+ v^+ +m_b q^+ -2m_b v^+ v\cdot k)} \right] \left[ \delta(xp_B^+ -p_b^+ +q^+) - \delta(x p_B^+ -p_b^+) \right] \,,\nn\\
\end{eqnarray}
where $\tilde \mu = \mu (e^{\gamma_E}/4\pi)^{1/2}$ and $p_B^+ = m_B v^+$, $p_b^+ = m_b v^+ + k^+$.

In the peak region where $x\sim 1-\lambda$, $x p_B^+ -p_b^+ \sim \mathcal{O}(\Lambda_{\textrm{QCD}})$. The delta function constrains $q^+\sim \mathcal{O}(\Lambda_{\textrm{QCD}})$, which allows the integrand to be expanded and simplified in terms of $q^+$,
\begin{eqnarray}
  && \left. S^{\textrm{QCD}(a)}(x,\mu) \right|_{\textrm{peak}} = \int_0^{p_b^+} dq^+ \frac{2}{q^+} \left[ \frac{1}{\epsilon_{\textrm{UV}}} + \frac{v^{+2}\mu^2}{q^+(q^+ -2v^+ v\cdot k)} \right] \left[ \delta(xp_B^+ -p_b^+ +q^+) - \delta(x p_B^+ -p_b^+) \right] \,. \nn\\
\end{eqnarray}
Further applying the plus distribution defined in Eq.\,(\ref{def_plus}), the amplitude simplifies to:
\begin{eqnarray}
  && \left. S^{\textrm{QCD}(a)}(x,\mu) \right|_{\textrm{peak}} = \bigg[ \frac{2}{m_b v^+ \! -x m_B v^+ \! +k^+} \, \left( \frac{1}{\epsilon_{\textrm{UV}}} +\ln\frac{v^{+2}\mu^2}{(m_b v^+ \! -x m_B v^+ \! +k^+)(m_b v^+ \! -x m_B v^+ \! +k^+ \! -2v^+ v\cdot k)} \right)  \nn\\
  && \qquad \times \theta(m_b v^+ \! -x m_B v^+ \! +k^+)\bigg]_\oplus \nn\\
  && \qquad +\bigg[ \int_{xm_B v^+ + k^+}^{\Lambda} dt \, \frac{2}{m_b v^+ \! -xm_B v^+ \! +t} \,\left( \frac{1}{\epsilon_{\textrm{UV}}} +\ln\frac{v^{+2}\mu^2}{(m_b v^+ \! -xm_B v^+ \! +t)(m_b v^+ \! -xm_B v^+ \! +t -2v^+ v\cdot k)} \right) \bigg] \nn\\
  && \qquad \times \delta(x m_B v^+ \! -m_b v^+ \! -k^+) \,.
\end{eqnarray}
After applying the $\overline{\textrm{MS}}$ renormalization scheme, the above expression becomes Eq.\,(\ref{SQCDa}a).

In the tail region where $x\sim 0$, the second delta function in Eq.\,(\ref{ampQCDa}) does not contribute and $x p_B^+ -p_b^+ \sim \mathcal{O}(m_b)$. The amplitude now takes the form
\begin{eqnarray}\label{tailfordiscussion}
  && \left. S^{\textrm{QCD}(a)}(x,\mu) \right|_{\textrm{tail}} = \frac{xm_B/m_b}{m_b v^+ -xm_B v^+} \left[ \frac{1}{\epsilon_{\textrm{UV}}} + \ln\frac{v^{+2}\mu^2}{(m_b v^+ - x m_B v^+)^2} \right] \theta(m_b v^+ -xm_B v^+) \,.
\end{eqnarray}
We set the heavy meson mass equal to the heavy quark mass for simplicity, since the difference is a power correction beyond the leading-power accuracy of the treatment. Therefore, the amplitude reduces to:
\begin{eqnarray}
  && \left. S^{\textrm{QCD}(a)}(x,\mu) \right|_{\textrm{tail}} = \frac{1}{m_b v^+} \frac{2x}{1-x} \left[ \frac{1}{\epsilon_{\textrm{UV}}} + \ln\frac{\mu^2}{(1-x)^2 m_b^2} \right] \theta(1-x) \,.
\end{eqnarray}
After applying the $\overline{\textrm{MS}}$ renormalization scheme, the above expression becomes Eq.\,(\ref{SQCDa}b).

For the box graph in Fig.\,\ref{All-in-QCD}(b). The corresponding amplitude is:
\begin{eqnarray}\label{ampQCDb}
  && S^{\textrm{QCD}(b)}(x,\mu) = \frac{2\pi}{\alpha_s C_F} \int \frac{d^d q}{(2\pi)^d} \bar u(p_b) (ig_s t^a \gamma_\mu \tilde\mu^\epsilon) \frac{i(\slash{p}_b -\slash{q} +m_b)}{(p_b-q)^2-m_b^2 +i\epsilon} \frac{-i}{q^2+ i\epsilon} \frac{\slash{n}_+}{2} \frac{i(\slash{p}_b -\slash{q}+m_b)}{(p_b-q)^2 -m_b^2 +i\epsilon} (i g_s t^a \gamma^\mu \tilde\mu^{\epsilon}) u(p_b) \nn\\
  &&\qquad \times \frac{1}{p_b^+} \delta(xp_B^+ -p_b^+ +q^+) \nn\\
  &&\qquad  = \int_{0}^{p_b^+} d q^+ \left[ \frac{q^+}{p_b^{+2}}\frac{1}{\epsilon_{\textrm{UV}}} -\frac{2q^+}{p_b^{+2}} -\frac{v^+(m_b^2(3p_b^+ -4q^+)-p_b^2(p_b^+-2q^+)-2p_b^- q^{+2})}{p_b^{+2}(2k^- q^+ v^+ + m_b(q^+-2v^+ v\cdot k))} \right. \nn\\
  &&\qquad + \left. \frac{q^+}{p_b^{+2}} \ln\frac{v^+(m_b v^+ +k^+)\mu^2}{q^+ (m_b q^+ +2k^- q^+ v^+ -2m_b v^+ v\cdot k)} \right] \delta(xp_B^+ -p_b^+ +q^+) \,.
\end{eqnarray}
In peak region, we have $x p_B^+ -p_b^+ \sim \mathcal{O}(\Lambda_{\textrm{QCD}})$. The delta function constrains $q^+\sim \mathcal{O}(\Lambda_{\textrm{QCD}})$, thus this amplitude can be expanded and simplified to:
\begin{eqnarray}
  && \left. S^{\textrm{QCD}(b)}(x,\mu) \right|_{\textrm{peak}} = \left[ \frac{m_b v^+ - xm_B v^+ +k^+}{m_b^2 v^{+2}}\frac{1}{\epsilon_{\textrm{UV}}} -\frac{2}{k^+ +m_b v^+ -xm_B v^+ -2v^+ v\cdot k} \theta(m_b v^+ -xm_B v^+ +k^+) \right]_\oplus \nn\\
  && \qquad + \left[ \int_{xm_B v^+ -m_b v^+}^{\Lambda} \!\! dt \left(  \frac{m_b v^+ - xm_B v^+ +t}{m_b^2 v^{+2}}\frac{1}{\epsilon_{\textrm{UV}}}-\frac{2}{t+m_b v^+ -xm_B v^+ -2v^+ v\cdot k} \right) \right] \delta(k^+ +m_b v^+ -xm_B v^+) \,.\nn\\
\end{eqnarray}
Upon the $\overline{\textrm{MS}}$ renormalization, the preceding expression simplifies to Eq.\,(\ref{SQCDb}a).

In the tail region where $x\sim 0$, this amplitude takes the form:
\begin{eqnarray}
  && \left. S^{\textrm{QCD}(b)}(x,\mu) \right|_{\textrm{tail}} \!\!=\!\! \left[ \frac{m_b v^+ -xm_B v^+}{m_b^2 v^{+2}}\frac{1}{\epsilon_{\textrm{UV}}} \!-\! \frac{m_b^2 +x^2 m_B^2 -(m_b -xm_B)^2\ln\frac{\mu^2}{(m_b-xm_B)^2} }{m_b^2 v^+(m_b -xm_B)} \right] \theta(m_b v^+ -xm_B v^+) .
\end{eqnarray}
Take $m_B \to m_b$, their difference belongs to high power correction,
\begin{eqnarray}
  && \left. S^{\textrm{QCD}(b)}(x,\mu) \right|_{\textrm{tail}} = \frac{1}{m_b v^+} \left[ (1-x)\frac{1}{\epsilon_{\textrm{UV}}} -\frac{1+x^2}{1-x} +(1-x)\ln\frac{\mu^2}{(1-x)^2 m_b^2} \right] \theta(1-x) \,.
\end{eqnarray}
This expression would change to Eq.\,(\ref{SQCDb}b) after renormalization.

In the case of local vertex graph in Fig.\,\ref{All-in-QCD}(c), we write down the amplitude according to Feynman rules:
\begin{eqnarray}\label{ampQCDa}
  && S^{\textrm{QCD}(c)}(\mu) = \frac{2\pi}{\alpha_s C_F} \int \frac{d^d q}{(2\pi)^d} \bar u(p_b) (ig_s t^a \gamma_\mu \tilde\mu^\epsilon) \frac{i(\slash{p}_b -\slash{q} +m_b)}{(p_b-q)^2 -m_b^2 +i\epsilon} \frac{-i}{q^2 +i\epsilon} \frac{\slash{n}_+}{2} \frac{i(\slash{p}_b -\slash{q} +m_b)}{(p_b-q)^2 -m_b^2 +i\epsilon} (ig_s t^a \gamma^\mu \tilde\mu^\epsilon) u(p_b) \frac{1}{p_b^+} \nn\\
\end{eqnarray}
As previously discussed in the context above Eq.\,(\ref{tailfordiscussion}), this local graph makes no contribution in the tail region. The computation of this amplitude in the peak region is straightforward:
\begin{eqnarray}
  && \left. S^{\textrm{QCD}(c)}(\mu) \right|_{\textrm{peak}} = \frac{1}{2\epsilon_{\textrm{UV}}}+2-\ln\frac{m_b^3}{(-2v\!\cdot\!k)^2\mu}  \,.
\end{eqnarray}
After renormalization, this expression reduces to Eq.\,(\ref{SQCDc}a).

%%%%%%%%%%%%%%%%%%%%%%%
\subsection{Shape function $S^{\textrm{HQET}}(\omega,\mu)$ defined in HQET}
%%%%%%%%%%%%%%%%%%%%%%%
\begin{figure}[htbp]
  \centering
  \includegraphics[width=0.75\columnwidth]{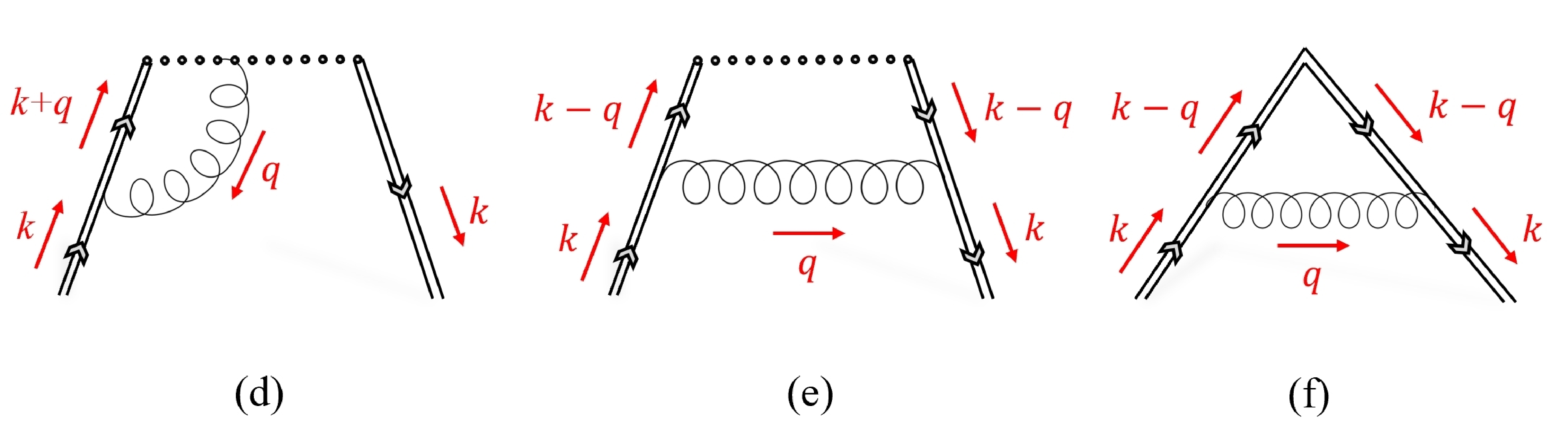}
  \caption{Feynman diagrams for the $B$-meson shape function defined in HQET. The relevant momentum notions are labeled.}
  \label{All-in-HQET}
\end{figure}
We begin with the heavy-quark sail graph, where the relevant momentum labels are specified in Fig.\,\ref{All-in-HQET}(d). The corresponding amplitude reads:
\begin{eqnarray}\label{ampHQETd}
  && S^{\textrm{HQET}(d)}(\omega,\mu) = \frac{2\pi}{\alpha_s C_F} \int \frac{d^d q}{(2\pi)^d} \frac{i}{-q^+ +i\epsilon} (ig_s t^a n_+^\mu \tilde\mu^{\epsilon}) \frac{-i}{q^2 +i\epsilon} \, \frac{i}{v\!\cdot\!(q+k) +i\epsilon} (ig_s t^a v_\mu \tilde\mu^\epsilon) \nn\\
  &&\qquad \times 2\left[\delta (\omega v^+ -q^+ -k^+) - \delta(\omega v^+ -k^+)\right] \nn\\
  &&\qquad = \left[ \left( \frac{2}{\epsilon_{\textrm{UV}}} \frac{1}{k^+ -\omega v^+} + \frac{2}{k^+ -\omega v^+} \ln\frac{v^{+2}\mu^2}{(k^+ -\omega v^+)(k^+ -\omega v^+ -2v^+ v\!\cdot\!k)} \right) \theta(k^+ -\omega v^+) \right]_\oplus \nn\\
  &&\qquad -\bigg[ \frac{1}{\epsilon_{\textrm{UV}}^2} +\frac{1}{\epsilon_{\textrm{UV}}}\ln\frac{\mu^2}{\omega^2} +\frac{2}{\epsilon_{\textrm{UV}}}\int_{\Lambda}^{2\omega v^+} \frac{dt}{t-\omega v^+} +\frac{\pi^2}{12} +\frac{1}{2}\ln^2\frac{\omega^2}{\mu^2} \nn\\
   &&\qquad -\int_{2\omega v^+}^{\Lambda} dt\, \frac{2}{t-\omega v^+} \ln\frac{v^{+2}\mu^2}{(t -\omega v^+)(t -\omega v^+ -2v^+ v\!\cdot\!k)} \bigg] \delta(\omega v^+ -k^+) \,.
\end{eqnarray}
The observed $1/\epsilon_{\textrm{UV}}^2$ pole emerges owing to the coexistence of UV and collinear divergences in the frame of HQET (cusp divergence). Another essential feature is that Eq.\,(\ref{ampHQETd}) does not have the restriction $\omega v^+ >0$, so the theta function is satisfied for $-\infty<\omega v^+ <k^+$, rather than $0<\omega v^+ <k^+$. The above expression will simplify to Eq.\,(\ref{HQETd}) through $\overline{\textrm{MS}}$ renormalization.

The amplitude of box graph in Fig.\,\ref{All-in-HQET}(e) is
\begin{eqnarray}\label{ampHQETd}
  && S^{\textrm{HQET}(e)}(\omega,\mu) = \frac{2\pi}{\alpha_s C_F} \int \frac{d^d q}{(2\pi)^d} (i g_s t^a v^\mu \tilde\mu^\epsilon) \frac{i}{v\!\cdot\!(k-q) +i\epsilon} \, \frac{-i}{q^2+i\epsilon}(ig_s t^a \tilde\mu^\epsilon v_\mu) \frac{i}{v\!\cdot\!(k-q) +i\epsilon} \delta(\omega v^+ -k^+ +q^+) \nn\\
  &&\qquad = -\frac{g_s^2 C_F}{4\pi^2} e^{\gamma_E \epsilon} \mu^{2\epsilon} \Gamma(1+\epsilon) \int_0^\infty dq^+  \frac{(v^+)^{2\epsilon}}{q^+ -2v^+ v\!\cdot\!k} \frac{1}{[q^{+2}-2q^+ v^+ v\!\cdot\!k]^\epsilon} \delta(\omega v^+ -k^+ +q^+) \,.
\end{eqnarray}
The integration over $q^+$ is straightforward, and as one can see, this graph does not contain ultraviolet divergence. Further applying the plus distribution, we have
\begin{eqnarray}
  && S^{\textrm{HQET}(e)}(\omega,\mu) = -\left[ \frac{2}{k^+ - \omega v^+ -2v^+ v\!\cdot \!k} \theta(k^+ -\omega v^+ ) \right]_\oplus  -\left[ \int_{\omega v^+}^{\Lambda} dt \, \frac{2}{t-\omega v^+ -2 v^+ v\!\cdot \! k} \right] \delta(\omega v^+ -k^+) \,.
\end{eqnarray}
This is the result of Eq.\,(\ref{HQETe}).

Finally, we arrive at the calculation of the local vertex graph in Fig.\,\ref{All-in-HQET}(f),
\begin{eqnarray}
  && S^{\textrm{HQET}(f)}(\mu) = \frac{2\pi}{\alpha_s C_F} \int \frac{d^d q}{(2\pi)^d} (i g_s t^a v^\mu \tilde\mu^\epsilon) \frac{i}{v\!\cdot\!(k-q)+i\epsilon} \, \frac{-i}{q^2 +i\epsilon} (ig_s t^a v_\mu \tilde\mu^\epsilon) \frac{i}{v\!\cdot\!(k-q)+i\epsilon} \nn\\
  &&\qquad = -2 e^{\gamma_E \epsilon} \mu^{2\epsilon} \Gamma(1+\epsilon) (v^+)^{2\epsilon} \int_0^\infty d q^+ \frac{1}{\left[ q^+ -2v^+ v\!\cdot\!k \right]^{1+\epsilon}} \, \frac{1}{q^{+\epsilon}} \nn\\
  &&\qquad = - \left( \frac{1}{\epsilon_{\textrm{UV}}} +\ln\frac{\mu^2}{(-2 v\!\cdot\! k)^2} \right) \,.
\end{eqnarray}
The renormalized result is presented in Eq.\,(\ref{HQETf}).

\end{widetext}

\end{document}